\documentclass[11pt]{article}
\usepackage{fancyhdr}
 \usepackage{graphicx}
\usepackage{amsmath}
\usepackage{shuffle}
\usepackage{amssymb}
\usepackage{wasysym}
\usepackage{color}
\usepackage{hyperref}
\def\Z{{\mathchoice {\hbox{$\sf\textstyle Z\kern-0.4em Z$}}
{\hbox{$\sf\textstyle Z\kern-0.4em Z$}}
{\hbox{$\sf\scriptstyle Z\kern-0.3em Z$}}
{\hbox{$\sf\scriptscriptstyle Z\kern-0.2em Z$}}}}
\def\N{{\mathchoice {\hbox{$\sf\textstyle N\kern-0.4em N$}}
{\hbox{$\sf\textstyle I\kern-0.42em N$}}
{\hbox{$\sf\scriptstyle I\kern-0.2em N$}}
{\hbox{$\sf\scriptscriptstyle I\kern-0.em N$}}}}
\def\nbR{\ensuremath{\mathrm{I\!R}}} 
\def\nbN{\ensuremath{\mathrm{I\!N}}} 
\def\nbC{{\mathchoice {\setbox0=\hbox{$\displaystyle\rm C$}%
\hbox{\hbox to0pt{\kern0.4\wd0\vrule height0.9\ht0\hss}\box0}}
{\setbox0=\hbox{$\textstyle\rm C$}\hbox{\hbox
to0pt{\kern0.4\wd0\vrule height0.9\ht0\hss}\box0}}
{\setbox0=\hbox{$\scriptstyle\rm C$}\hbox{\hbox
to0pt{\kern0.4\wd0\vrule height0.9\ht0\hss}\box0}}
{\setbox0=\hbox{$\scriptscriptstyle\rm C$}\hbox{\hbox
to0pt{\kern0.4\wd0\vrule height0.9\ht0\hss}\box0}}}}
\def\nbQ{{\mathchoice {\setbox0=\hbox{$\displaystyle\rm
Q$}\hbox{\raise
0.15\ht0\hbox to0pt{\kern0.4\wd0\vrule height0.8\ht0\hss}\box0}}
{\setbox0=\hbox{$\textstyle\rm Q$}\hbox{\raise
0.15\ht0\hbox to0pt{\kern0.4\wd0\vrule height0.8\ht0\hss}\box0}}
{\setbox0=\hbox{$\scriptstyle\rm Q$}\hbox{\raise
0.15\ht0\hbox to0pt{\kern0.4\wd0\vrule height0.7\ht0\hss}\box0}}
{\setbox0=\hbox{$\scriptscriptstyle\rm Q$}\hbox{\raise
0.15\ht0\hbox to0pt{\kern0.4\wd0\vrule height0.7\ht0\hss}\box0}}}}
\def\nbT{{\mathchoice {\setbox0=\hbox{$\displaystyle\rm
T$}\hbox{\hbox to0pt{\kern0.3\wd0\vrule height0.9\ht0\hss}\box0}}
{\setbox0=\hbox{$\textstyle\rm T$}\hbox{\hbox
to0pt{\kern0.3\wd0\vrule height0.9\ht0\hss}\box0}}
{\setbox0=\hbox{$\scriptstyle\rm T$}\hbox{\hbox
to0pt{\kern0.3\wd0\vrule height0.9\ht0\hss}\box0}}
{\setbox0=\hbox{$\scriptscriptstyle\rm T$}\hbox{\hbox
to0pt{\kern0.3\wd0\vrule height0.9\ht0\hss}\box0}}}}
\def\nbS{{\mathchoice
{\setbox0=\hbox{$\displaystyle     \rm S$}\hbox{\raise0.5\ht0%
\hbox to0pt{\kern0.35\wd0\vrule height0.45\ht0\hss}\hbox
to0pt{\kern0.55\wd0\vrule height0.5\ht0\hss}\box0}}
{\setbox0=\hbox{$\textstyle        \rm S$}\hbox{\raise0.5\ht0%
\hbox to0pt{\kern0.35\wd0\vrule height0.45\ht0\hss}\hbox
to0pt{\kern0.55\wd0\vrule height0.5\ht0\hss}\box0}}
{\setbox0=\hbox{$\scriptstyle      \rm S$}\hbox{\raise0.5\ht0%
\hboxto0pt{\kern0.35\wd0\vrule height0.45\ht0\hss}\raise0.05\ht0%
\hbox to0pt{\kern0.5\wd0\vrule height0.45\ht0\hss}\box0}}
{\setbox0=\hbox{$\scriptscriptstyle\rm S$}\hbox{\raise0.5\ht0%
\hboxto0pt{\kern0.4\wd0\vrule height0.45\ht0\hss}\raise0.05\ht0%
\hbox to0pt{\kern0.55\wd0\vrule height0.45\ht0\hss}\box0}}}}
\def\nbZ{{\mathchoice {\hbox{$\sf\textstyle Z\kern-0.4em Z$}}
{\hbox{$\sf\textstyle Z\kern-0.4em Z$}}
{\hbox{$\sf\scriptstyle Z\kern-0.3em Z$}}
{\hbox{$\sf\scriptscriptstyle Z\kern-0.2em Z$}}}}

\def\nbN{{\mathchoice {\hbox{$\sf\textstyle N\kern-0.4em N$}}
{\hbox{$\sf\textstyle I\kern-0.4em N$}}
{\hbox{$\sf\scriptstyle I\kern-0.2em N$}}
{\hbox{$\sf\scriptscriptstyle I\kern-0.1em N$}}}}
\def\P{{\mathchoice {\hbox{$\sf\textstyle P\kern-0.4em Z$}}
{\hbox{$\sf\textstyle P\kern-0.4em P$}}
{\hbox{$\sf\scriptstyle P\kern-0.3em P$}}
{\hbox{$\sf\scriptscriptstyle P\kern-0.2em P$}}}}
\def\cqfd{\par\nopagebreak\rightline{\vrule height 3pt width 5pt depth 2pt}
\def\R{{\mathchoice {\hbox{$\sf\textstyle R\kern-0.4em R$}}
{\hbox{$\sf\textstyle I\kern-0.42em R$}}
{\hbox{$\sf\scriptstyle I\kern-0.2em R$}}
{\hbox{$\sf\scriptscriptstyle I\kern-0.1em R$}}}}
\medbreak}
\def\nbQ{{\mathchoice {\setbox0=\hbox{$\displaystyle\rm
Q$}\hbox{\raise
0.15\ht0\hbox to0pt{\kern0.4\wd0\vrule height0.8\ht0\hss}\box0}}
{\setbox0=\hbox{$\textstyle\rm Q$}\hbox{\raise
0.15\ht0\hbox to0pt{\kern0.4\wd0\vrule height0.8\ht0\hss}\box0}}
{\setbox0=\hbox{$\scriptstyle\rm Q$}\hbox{\raise
0.15\ht0\hbox to0pt{\kern0.4\wd0\vrule height0.7\ht0\hss}\box0}}
{\setbox0=\hbox{$\scriptscriptstyle\rm Q$}\hbox{\raise
0.15\ht0\hbox to0pt{\kern0.4\wd0\vrule height0.7\ht0\hss}\box0}}}}
\usepackage{fancyhdr}
\title{\bf Ten Conferences {\it WORDS: }Open Problems and Conjectures}
\author{Jean N\'eraud
}
\date{}
\renewcommand{\headrulewidth}{0pt}
\begin{document}
\thispagestyle{empty}
\maketitle
\thanks{68- Laboratoire d'Informatique, de Traitement de l'Information et des Syst\`emes
Normandie Univ, UNIROUEN, UNIHAVRE, INSA Rouen
 LITIS, 76000 Rouen, France}
\thanks{
\begin{center}
{jean.neraud@univ-rouen.fr, neraud.jean@free.fr}
\end{center}
\ \\

\abstract{In connection to the development of the field of Combinatorics on Words, we present a list of open problems and conjectures that were stated during the ten last meetings WORDS.
We wish to continually update the present document by adding informations concerning advances in problems solving. }
\newpage
\ \\
\ \\
\tableofcontents

\newpage

%
%
\section*{\it Foreword}
\addcontentsline{toc}{section}{\it Foreword}

The first conference WORDS was organized in 1997 in Rouen, France. Since then, a series of ten meetings held. In \cite{N2015,N2016}, we provided a summary of the contributions which were presented in connection with the development of the field of Combinatorics on Words.\\

The aim of the present note is to bring complementary informations, with two key objectives:

- Beforehand,  we provide a nomenclature of some of the challenging conjectures and problems which were stated during these conferences.

- With regard to the the state-of-the-field, we hardly wish to continually update the present  study by including the most recent advances in the framework of the listed questions.\\

From a practical point of view, each of these questions or conjectures is nomenclatured by referencing to its topic and to the meeting WORDS were it was stated, with bibliographic references. A short introduction to the problematic is also given.
The present document should be gradually updated:
in view of this, please contact its author, bibliographic references being clearly welcome.\\

Evidently, each of the numerous results, questions and conjectures which were presented during the ten conferences WORDS plays a noticeable part in the state-of-the-field. From this point of view, we wish that our study will bring valuable information to researchers of the community

\section*{\it 1. The topic of patterns}
\addcontentsline{toc}{section}{\it 1. The topic of patterns}
\thispagestyle{empty}
 \renewcommand{\headrulewidth}{0px} 
Let $\Sigma$, $A$ be two finite alphabets and let $p\in\Sigma^*$, $w\in A^*\cup A^\omega$.  We say that the word $w$ {\it encounters}  $p$ if a non-erasing morphism $h:\Sigma^*\longrightarrow A^*$ exists such that $w\in A^*h(p)(A^*\cup A^\omega)$;
otherwise the word $w$ {\it avoids $p$} or equivalently said is $p$-free. In this context $p$ is refered as  a {\it pattern}, moreover we impose that the morphism $h$ satisfies $h(a)=a$ for any letter $a\in\Sigma\cap A$.
The pattern $p$ is $k$-{\it avoidable} if an infinite word avoiding $p$ exists over a $k$-letter alphabet.
From this point of view, it is well known that the infinite word of Thue-Morse have the fundamental property that it avoids any pattern of type $aXaXa$, with 
 $a\in A$ and $X\in \Sigma^*$
\subsection*{\it 1.1 Avoidance of patterns}
\addcontentsline{toc}{subsection}{\it 1. 1.1 Avoidance of patterns}
Avoidance of patterns is a central question in the topic and has inspired lots of questions: \\
\ \\
{\it --WORDS 1997:} \\
{\it Authors:}  Roman Kolpakov, Gregory Kucherov and Yuri Tarannikov \cite[pp. 161--175]{Issue1997}.\\
For a natural $n\geq 2$, a word is $n${\it th power-free} if it does not contain  a  $n$th power of a non-empty word as a factor. 
Given $A=\{0,1\}$, denote by {\it PF(n)} the corresponding set of such words and set  $\rho(n)={\rm{\underline l}{\underline i}{\underline m}}~ _{k\rightarrow \infty}(\frac{1}{k} \cdot{\rm min}\{|w|_1: w\in{\it PF(n)}\cap A^k\}) $ (the minimal density of the letter $1$ in the words of $PF(n)$).\\
In their paper, the authors prove that: $$(\forall n\geq 3)~(\exists C>0)~~\rho(n)\leq\frac{1}{n}+\frac{1}{n^3}+\frac{1}{n^4}+\frac{C}{n^5}$$
$\rho$ can be extended to real arguments: given a real $x\in \nbR$,
denote by  {\it PF}$(x)$ the set of the binary words that do not contain a factor of exponent greater than or equal to $x$.
The authors proved that $\rho$ is discontinuous to the right in each point of $\{7/3\}\cup \{n\in\nbN:n\geq 3\}$, moreover, they asked the following questions:
%
\begin{itemize}
\item
\color{blue} {\it Question 1.1.97.1:} \color{black} Does $\rho$ has other discontinuity? What  are they? Is $\rho$ piece-wise constant?

\item
\color{blue} {\it Question 1.1.97.2: }\color{black} If  a pattern is not $k$-avoidable, but is $(k+1)$-avoidable, what is the minimal frequency of a letter in an infinite word over $k+1$ letters that avoids that pattern?

\item
\color{blue} {\it Question 1.1.97.3:} \color{black}  Kirby Baker, Georges Mac Nulty and Walter Taylor have shown that the pattern $abXbcYcaZ baTac$ is $4$-avoidable,  but not $3$-avoidable \cite{BMT89}.
 What is the minimal proportion of the fourth letter needed to avoid that pattern?
\end{itemize}
\ \\
{\it --WORDS 2003:}\\
{\it Author:} James Currie \cite[pp. 7--18]{Issue2003}.\\
The author reviews results concerning of words avoiding pattern. He recall a lot of open problems. Let's begin
by two purely algorithmic questions:
\begin{itemize}
\item
\color{blue} {\it Question 1.1.03.1:} \color{black} Is it decidable, given a pattern $p$ and an integer $k$, whether $p$ is $k$-avoidable?
\item
\color{blue} {\it Question 1.1.03.2:} \color{black} Given  a pattern $p$, what is the complexity of deciding whether $p$ is avoidable? 
\end{itemize}
With regard to t$k$-avoidability itself, three open problems are stated:
\begin{itemize}
\item
\color{blue} {\it Question 1.1.03.3:}  \color{black} Is there a patten that is $6$-avoidable but not $5$-avoidable?
\item
\color{blue} {\it Question 1.1.03.4:} \color{black} Is $aabaacbaab$ $3$-D0L-avoidable (i.e. is there a ternary morphism $g$ such that $g^\omega(a)$ avoids $aabaacbaab$)?
\item
\color{blue} {\it Conjecture 1.1.03.5:} \color{black} If a pattern is $k$-avoidable then it is $k$-HD0L-avoidable (i.e. are there morphisms $f:\Sigma^*\longrightarrow A^*$, $g:\Sigma^*\longrightarrow\Sigma^*$, with $|\Sigma|=k$ such that  $f(g^\omega(a))$ avoids $p$)?
\end{itemize}
The so-called probabilistic method is often use in tackling many problems in discrete mathematics \cite[pp. 3--5]{AS2016}.
Trying to prove that a
structure with certain properties exists, this method consists in constructing a convenient probability
space of structures and then shows that the desired properties hold in this space with
a non-zero probability.
\begin{itemize}
\item
\color{blue} {\it Question 1.1.03.6:} \color{black} Explore the applications of  the probabilistic method in the scope of pattern avoidance.
\end{itemize}
Circular words are concerned by the following questions:
\begin{itemize}
\item
\color{blue} {\it Conjecture 1.1.03.7: } \color{black} If $p$ is $k$-avoidable, then arbitrary long $k$-letter circular words avoiding $p$ exist.
\item
\color{blue} {\it Conjecture 1.1.03.8:} \color{black} If $p$ is $k$-avoidable that a $k$-letter circular word of length $|p|$ avoiding $p$ exist.
\item
\color{blue} {\it Conjecture 1.1.03.9:} \color{black} Let $p$ be $k$-avoidable.
\begin{enumerate}
\item
If the number of $p$-free words on $k$ letters of length $n$ grows exponentially with $n$, then an integer $N_0$ exists such that, for every $n>N_0$,  there are circular $p$-free words on $k$ letters with length $n$.
\item
If the number of $p$-free words on $k$ letters of length $n$ grows polynomially with $n$, then
the set of possible lengths for circular $p$-free words on $k$ letters
has density $0$ in the set $\nbN\setminus\{0\}$.
\end{enumerate}
\item
\color{blue} {\it Question  1.1.03.10:} \color{black} The number of $k$-power-free binary words of length $n$ grows polynomially with $n$ for $k\leq 7/3$, but exponentially for $k>7/3$ \cite{KS04}. Examine analogous results for alphabets of arbitrary size.
\item
\color{blue} {\it Conjecture 1.1.03.11:} \color{black} Extension of a result from \cite{BMT89}: the set of circular words over $\{0,1,2,3\}$ avoiding the pattern $abXbcYcaZbaTcb$ has density $0$ in the set $\nbN\setminus\{0\}$.
\end{itemize}
A word $w$ is maximal $p$-free if $p$ encounters any word in $\Sigma w\Sigma$. The three following conjectures are stated:
\begin{itemize}
\item
\color{blue} {\it Conjecture 1.1.03.12:} \color{black} Let $\Sigma$ be  an alphabet and let $w\in\Sigma^*$ a $p$-free word. Then $w$ is a factor of a maximal $p$-free word over $\Sigma$.
\item
\color{blue} {\it Conjecture 1.1.03.13:} \color{black} Given an alphabet $\Sigma$ and a pattern $p$, maximal $p$-free word over $\Sigma$ exists.
\item
\color{blue} {\it Conjecture 1.1.03.14:} \color{black} Let  $\Sigma$ be an alphabet, $k\in [1,2]$, and $w\in\Sigma^*$ be a $k$-power-free word. Then in any case, $w$ is a factor of a maximal $k$-power-free word over $\Sigma$.
\end{itemize}
\ \\
\ \\
{\it --WORDS 2007:}\\
{\it Authors: }Inna Mikhailova and Mikhail Volkov \cite[pp. 212--221]{Acts2007}.\\
The authors prove that every avoidable pattern can be avoided by an infinite sequence of palindromes over a fixed alphabet.
\begin{itemize}
\item
\color{blue} {\it Question 1.1.07.1:} \color{black} 
Is it possible to avoid an arbitrary pattern $p$ by an infinite sequence of palindromes over each alphabet on which $p$ is avoidable?
\end{itemize}
\ \\
{\it --WORDS 2011:}\\
{\it Authors: }Helena Petrova and Arseny Shur  \cite[pp. 1595-1611]{Issue2011}.\\
 With respect to the prefix (suffix) order, any repetition-free language can be viewed as a poset whose diagram is a tree, each node generating a subtree and
being a common prefix (suffix) of its descendants. The authors asked the three following questions. In fact it has been shown in  \cite{C95}that the first one is decidable for some power-free languages:
\begin{itemize}
\item
\color{blue} {\it Question 1.1.11.1: }\color{black}
Does a given word generate a finite or infinite subtree?
\end{itemize}
In the case of a single word,  in \cite{BEMN79} it is shown  that for all
$k$-th power-free
languages, the subtree generated by any word has at least one leaf.
\begin{itemize}
\item
\color{blue} {\it Question 1.1.11.2:} \color{black}
Are the subtrees generated by two given words isomorphic?
\end{itemize}
The authors prove that in the langage of cube-free words arbitrarily large finite subtrees may be generated.
\begin{itemize}
\item
\color{blue} {\it Question 1.1.11.3 (generalization of  \cite[Problem 1.10]{AS03} to arbitrary words): } \color{black} 
 Can words generate arbitrarily large finite subtrees?
\end{itemize}
\ \\
{\it --WORDS 2013}\\
{\it Authors: }Tero Harju, Mike M\"uller \cite[pp. 29--38]{Issue2013}.\\
Let $u_0,u_1$ be two words over an alphabet $A$, and let $\beta\in\{0,1\}^*$ with $|\beta|=|u_0|+|u_1|$, called the {\it conduction sequence}, such that $|\beta|_i=|u_i|$ ($i=0,1$). 
The {\it shuffle} of $u_0$ and $u_1$ conducted by $\beta$ is the word  $u_0\shuffle_\beta u_1$ whose letter of index $i$
($i\in [1,|u_0|+|u_1|]$) is  $u_{\beta(i)}(j)$,
where $j={\rm card}\{k\in [1,i]~ |~ \beta(k) = \beta(i)\}$.  This definition can be extended to infinite words (one requires that $\beta$ contains infinitely many occurrences of both $0$ and $1$).
The authors proved that
an ternary infinite square-free word $u$ exists  such that $u$ can be shuffled with itself
to produce an infinite square-free word. They asked for the following questions:
\begin{itemize}
\item
\color{blue} {\it Question 1.1.13.1: } \color{black} Which square-free words $u$ can be shuffled to obtain a square-free word $u\shuffle_\beta u$?
\item
\color{blue} {\it Question 1.1.13.2: } \color{black} Which words $u$ can be shuffled to obtain a unique square-free word  $u\shuffle_\beta u$?\item
\color{blue} {\it Question 1.1.13.3:  }\color{black} Which words $w$ can be obtained in more than one way from a single word $u$ using different conducting sequences?
\item
\color{blue} {\it Question 1.1.13.4:  }\color{black} Which square-free words $w$ are themselves shuffles of square-free words: $w= u\shuffle u$?
\item
\color{blue} {\it Question 1.1.13.5 (due to I. Petrykiewicz): } \color{black} For any infinite ternary  square-free word $u$,
does  an infinite  ternary square-free word $w$ exists such that $u=u\shuffle_\beta w$ for some infinite $\beta$?
\item
\color{blue} {\it Question 1.1.13.6: }\color{black} Does an infinite square-free word $w$ exists such that $w=w\shuffle_\beta w$ for some infinite $\beta$?
\end{itemize}
\ \\
{\it --WORDS 2015:}\\
{\it Authors: }Helena Petrova and Arseny Shur \cite[pp. 223--236]{Acts2015}.\\
As mentionned above, the set of square-free words over a given alphabet may be represented by a prefix tree $T$ whose nodes are these square-free words. In WORDS 2015 the authors stated the following conjecture:
\begin{itemize}
\item
\color{blue} {\it Conjecture 1.1.15.1: }\color{black} In the tree $T$, the size of any minimal subtree of index $n$ is $O({\rm log}~ n)$.
\end{itemize}
\subsection*{\it 1.2 The repetition threshold}
\addcontentsline{toc}{subsection}{\it 1.2 The repetition threshold}
The {\it repetition threshold} for $k$ letters, 
which we denote by $RT(k)$, is the
 smallest rational number $\alpha$ such that there exists an infinite word whose finite factors have exponent at most $\alpha$. Actually, powers in of the Thue-Morse sequence have exponent at most $2$ and we have $RT(2) = 2$.\\
In the seventies, Fran\c{c}oise Dejean conjectured that for every $k>2$ the following holds:
$$
RT(k)=
\begin{cases}7/4~~{\rm if}~~k = 3\cr
7/5~~{\rm if}~~k = 4\cr
k/k-1~~{\rm otherwise.}\cr
\end{cases}
$$
Dejean's conjecture have been partially solved by different authors. The final proof was completed in 2009 by James Currie and Narad Rampersad for $15\leq n\leq 26$,
and independently by Micha\"el Rao for $8\leq k \leq 38$ \cite[pp. 3010--3018]{Issue2009}.\\
\ \\
{\it --WORDS 2005:}\\
{\it Author: }Pascal Ochem \cite[pp. 388--392]{Issue2005}.\\
A word is $\alpha$-free (resp.
$\alpha^+$-free) if it contains no factor that is an $\alpha'$-power, for any rational  $\alpha'\ge\alpha$ ($\alpha'>\alpha$).
\begin{itemize}
\item
\color{blue}{\it Question 1.2.05.1 (stronger version of Dejean's conjecture): }\color{black}
\begin{itemize} \item For every $k\geq 5$, an infinite $(k/k-1)^+$-free word over  $k$ letters  exists with letter frequency $1/k+1$.
\item For every $k\geq 6$, an infinite $(k/k-1)^+$-free word over  k-letter  exists with letter frequency $1/k-1$.
\end{itemize}
\newpage
\color{red} {\it Advances in problem solving} \color{black}
\begin{itemize}
\item
A partial solution for $9 \leq k \leq 38$  \cite[pp. 3010--3018]{Issue2009} was given by Rao.
\item The conjecture has been completely solved by Micha\"el Rao (private communication at WORDS 2015).
\end{itemize}
\end{itemize}
{--\it WORDS 2011}\\
{\it Authors: }
Golnaz Badkobeh and Maxime Crochemore \cite[pp. 37--43]{Acts2011}.\\
Starting with  $RT(k)$, the definition of {\it FRT}$(k)$, the {\it finite repetition threshold} for $k$ letters, stipulates that only a finite number of factors with exponent $\alpha$ may exist in the corresponding infinite word.
In 2008, Jeffrey Shallit proved that {\it FRT}$(2)=7/3$. In their presentation of WORDS 2011, Golnaz Badkobeh and Maxime Croche\-more proved that {\it FRT}$(3)=RT(3)=7/4$.
\begin{itemize}
\item
 \color{blue} {\it Conjecture 1.2.11.1: }\color{black}
We have {\it FRT}$(4)=RT(4)=7/5$.\\
\ \\
\color{red}
{\it Advances  in problem solving}
\color{black}\\
The conjecture was solved by Golnaz Badkobeh, Maxime Croche\-more and Micha\"el Rao. In addition they proved that {\it FRT}$(k)=RT(k)$ for $k\leq 6$ (private communication at WORDS 2015).
\end{itemize}
\subsection*{\it 1.3. On the number of different squares in a finite word}
\addcontentsline{toc}{subsection}{\it 1.3. On the number of different squares in a finite word}
\ \\
{\it --WORDS 2015}\\
{\it Authors: Florin Manea and Shinnosuke Seki} \cite[pp. 160--169]{Acts2015}.\\
 A natural question consists in examining the number of patterns that may appear in a finite word. 
From this point of view, Aviezri Fraenkel and Jamie Simpson focused to {\it dictinct}  squares defined as squares of different shape (not just translated of each other). At WORDS 1997, in the case of the sequence of Fibonacci words $(f_n)_{n\ge 0}$,  they showed that the exact number of such squares is $2(f_{n-2}-1)$, for any integer $n\ge 5$ \cite[pp. 95--106]{Issue1997}. In \cite{FS98} they proved that the number of distinct squares in an arbitrary word of length $n$ is bounded by $2n$. 
A refinement of 
$2n-O(\log n)$ was provided by Lucian Illie in WORDS 2005 \cite[pp. 373--376]{Issue2005} and the best bound known so far is due to $\frac{11n}{6}$ \cite{DFT15}.
\begin{itemize}
\item
\color{blue}
{\it Conjecture 1.3.05.1 (Square conjecture, due to A. Fraenkel and J. Simpson): }
\color{black}
The number of different squares in a word of length $n$ is bounded by $n$.
\end{itemize}
Define the square density of a word $w$ by $\rho_{\rm sq}(w)=\frac{\#\{x^2\in\Sigma^+|x^2~{\rm is~a~factor~of~}w\}}{|w|}$. In their contribution of WORDS 2015, the authors  proved that  binary words have the largest square density, and they asked the question of constructing a ``square-density" amplifier:
\begin{itemize}
\item
\color{blue}
{\it Question 1.3.15.1: }
\color{black}
 Can we compute a mapping $f:\Sigma^*\longrightarrow \Sigma^*$ for which a constant $c>1$ exists, such that for all $w\in\Sigma^*$, if $\rho_{\rm sq}(w) \ge 1$ then we have $\rho_{\rm sq}(f(w))\ge c\rho_{\rm sq}(w)$?
\end{itemize}
\subsection*{\it 1.4 The ``runs" conjecture}
\addcontentsline{toc}{subsection}{\it 1.4 The ``runs" conjecture}
A {\it run} may be defined as the occurrence of a  repetition of exponent at least $2$  that is maximal in the sense where it cannot be extended from left or right to obtain the same type pattern. Such objects play an important role in a lot of string matching algorithms.\\
\ \\
{\it --WORDS 2009}\\
 {\it Authors: }Maxime Crochemore, Lucian Ilie and Liviu Tinta \cite[2931--2941]{Issue2009}.\\
These authors showed that, given a word of length $n$, the number  of its runs is up-bounded by $1.029n$. This is a noticeable step in the proof of the so-called ``runs" conjecture:
\begin{itemize}
\item
\color{blue}
{\it Conjecture 1.4.09.1 (``runs" conjecture, due to Kolpakov and Kucherov, 1999): }
\color{black}
For a binary alphabet the number of runs is bounded by $n$.
\end{itemize}
\subsection*{\it 1.5 The prefix-suffix square completion}
\addcontentsline{toc}{subsection}{\it 1.5 The prefix-suffix square completion}
\ \\
{\it --WORDS 2015}\\
{\it Authors: }Marius Dumitran and Florin Manea \cite[147--159]{Acts2015}.\\
The so-called {\it suffix-square duplication} allows to derive from a word $w$ any word $wx$ such that $x$ is a suffix of $w$. The {\it suffix-square} completion, in turn, derives from a word $w$ a word $wx$ such that $w$ has a suffix  of $yxy$. {\it Prefix} and {\it prefix-square} duplication (completion) may be defined in a similar way. In their talk of WORDS 2015, Marius Dumitran and Florin Manea  made use of such operations for generating an infinite words that do not contains any repetition of exponent greater than $2$. With regards to combinatorics properties of words, they asked the following questions:
\begin{itemize}
\item
\color{blue}
{\it Question 1.5.15.1: }
\color{black}
What is the minimum exponent of a repetion which is avoidable by an infinite word constructed by iterated (prefix)-suffix duplication?
\item
\color{blue}
{\it Question 1.5.15.2: }
\color{black}
By applying prefix-suffix completion, can we construct  words that avoid cubes, and every word containing squares?
\item
\color{blue}
{\it Question 1.5.15.3: }
\color{black}
Does the language of finite words constructed, starting with a single word, by iterating prefix-suffix square completion  remains semi-linear?
\item
\color{blue}
{\it Question 1.5.15.4: }
\color{black}
Draw studies of languages constructed by iterating prefix-suffix square completion, starting with special sets of initial words such as singleton sets, finite sets, regular sets, etc.
\item
\color{blue}
{\it Question 1.5.15.5: }
\color{black}
What is the minimum number of steps of square completion needed to obtain a word from one of its factors?
\end{itemize}
\subsection*{\it 1.6 Abelian patterns}
\addcontentsline{toc}{subsection}{\it 1.6 Abelian patterns}
An {\it abelian square} consists in a pattern which is  obtained by applying a permutation on the letters of a square, say $u^2$. Clearly, with every pattern, a corresponding abelian one may be associated.
In 1992,  by constructing an {\it abelian square free} word over a four-letter alphabet, Veikko Ker\"anen solved a famous open problem formulated by Erd\"os in 1961 \cite{E61,K92}.  In WORDS 2007, he presented new abelian square-free morphisms and a powerful substitution over $4$ letters \cite[pp. 190--200]{Acts2007}.\\
\ \\
{\it --WORDS 2003}\\
{\it Author: } James Currie \cite[pp. 7--18]{Issue2003}.
\begin{itemize}
\item
{\it \color{blue}
Question 1.6.03.1: }
\color{black}
Which of the following patterns
$01020312$,
$01020321$,
$01021303$,
$01023031$,
$010203013$,
$010213020$
is avoidable in the abelian sense?
\item
{\it \color{blue}
Question 1.6.03.2: }
\color{black}
Show that the number of abelian cube-free ternary words grows exponentially with length.
\end{itemize}
 Given a $n$-letter alphabet, define the sequence $Z_n$ recursively by $Z_1=1,~~Z_n=Z_{n−1}nZ_{n−1},~n>1$.
\begin{itemize}
\item
{\it \color{blue}
Conjecture 1.6.03.3: }
\color{black}
Let $p $be any pattern over an alphabet of n letters. Then $p$ is abelian
avoidable iff $Z_n$ is $p$-free in the abelian sense.
\item
{\it \color{blue}
Question 1.6.03.4: }
\color{black}
Given pattern $p$ and integer $n$, what is the complexity of deciding whether $Z_n$
encounters p in the abelian sense?
\end{itemize}
Define the abelian repetitive threshold function  and the dual abelian repetitive threshold function on $(1, 2]$ by:\\
{\it ART}$(n) = \inf\{s : y^s$ is avoidable on n letters in the abelian sense$\}$\\
{\it DART}$(r) = \min\{n\in\N : y^r$ is avoidable in the abelian sense on n letter$\}$.
\begin{itemize}
\item
{\it \color{blue}
Question 1.6.03.5: }
\color{black}
What are the values of {\it ART}$(n)$ and {\it DART}$(r)$?
\end{itemize}
{\it --WORDS 2013}\\
Two papers were concerned by open questions:\\
\ \\
{\it Authors: }Mari Huova and Aleksi Saarela  \cite[161--168]{Acts2013}.\\
Two words $u$, $v$ are {\it $k$-abelian equivalents} if every word of length at most $k$ occurs as a factor in $u$ as many times as in $v$.
A word is a {\it strongly $k$-abelian $n$th-power} if it is $k$-abelian equivalent to a $n$th-power. In their contribution to WORDS 2013, the authors prove that
strongly $k$-abelian $n$th-powers are unavoidable on any alphabet, moreover they formulate the following questions:
\begin{itemize}
\item
{\it \color{blue}
Question 1.6.13.1: }
\color{black}
How many $k$-abelian equivalence classes of words of a given length contain an $n$th power?
\item
\color{blue}
Question 1.6.13.2:
\color{black}
How many words of a given length are strongly $k$-abelian $n$th powers?
\item
\color{blue}
{\it Question 1.6.13.3: }
\color{black}
What is the length of the longest word avoiding strongly $k$-abelian $n$th powers?
\item
\color{blue}
{\it Question 1.6.13.4: }
\color{black}
How many words avoid strongly $k$-abelian $n$th powers?
\item
\color{blue}
{\it Question 1.6.13.5: }
\color{black}
How many words of a given length contain a strongly $k$-abelian $n$th power?
\item
\color{blue}
{\it Question 1.6.13.6: }
\color{black}
How many words of a given length are strongly $k$-abelian $n$th powers?
\end{itemize}
{\it Author: }Micha\"el Rao \cite[pp. 39--46]{Issue2013}.\\
Given an integer $n\ge 2$, a word $u$ is a $k$-abelian-$n$-power if we have $u = u_1u_2\cdots u_n$, where $u_i$ and $u_{i+1}$ are $k$-abelian equivalents for every $i\in\{1,\cdots n-1\}$.
\begin{itemize}
\item
\color{blue}
{\it Question 1.6.13.7: }
\color{black}
Is there a pure morphic binary word avoiding $2$-abelian cubes?
\item
\color{blue}
{\it Question 1.6.13.8: }\\
\color{black}
(1) Can we avoid abelian-squares of the form $uv$, with $|u| \ge 2$, over a ternary alphabet?\\
(2) Can we avoid abelian-cubes of the form $uvw$, with $|u| \ge 2$, over a binary alphabet?
\item
\color{blue}
{\it Question 1.6.13.9: }
\color{black}
Is there a natural integer $p$ such that $2$-abelian-squares of period at least $p$ can be avoided over a binary alphabet?
\color{blue}
\item
{\it Question 1.6.13.9: }
\color{black} Is there a natural integer $p$ such that one can avoid abelian cubes of period at least $p$ over a binary alphabet?
\end{itemize}
The so-called additive powers consist in a generalization of abelian powers: given an alphabet $\Sigma\subseteq N$, an {\it additive $k$th power}  is a word $p_1\cdots p_k\in\Sigma^*$
such that $|p_1|=\cdots=|p_k|$, and $\sum (p_1)=\cdots=\sum(p_k)$, where $p_i$ stands for the sum of the digits of the word $p_i$ ($1\leq i\leq k$). 
In 2011 Cassaigne, Currie, Schaeffer and Shallit proved that additives cubes are avoidable on $\{0,1,2,3,4\}$ \cite{CCSS11}.
In WORDS 2013 Rao asked the following question:
\begin{itemize}
\item
\color{blue}
{\it Question 1.6.13.10: }\\
\color{black}
Are there infinite additive-cube-free words on the following alphabets: $\{0,1,2,3\}$, $\{0,1,4\}$ and $\{0,2,5\}$?
\end{itemize}
\ \\
{\it --WORDS 2015}\\
Open questions  were stated in two talks.\\
\ \\
{\it Authors: }Gabriele Fici and Filippo Mignosi \cite[pp. 122--134]{Acts2015}.\\
A word of length $n$ can contains $O(n^2)$ distinct abelian  squares \cite{KRRW2014}.
\begin{itemize}
\item
\color{blue}
{\it Conjecture 1.6.15.1: }
\color{black}
Assume that a word of length $n$ containing $k$ many distinct abelian-square factors exists.
Then a binary word of length $n$ containing at least $k$ many distinct abelian-square factors exists.
\end{itemize}
Two abelian squares are {\it inequivalent} if their Parikh vectors are different \cite{FSP97}.
\begin{itemize}
\item
\color{blue}
{\it Conjecture 1.6.15.2 (due to Kosciumaka, Radoszewski, Rytter, Wale\'n \cite{KRRW14}): }
\color{black}
A word of length $n$ contains $O(n\sqrt{n})$ inequivalent abelian-squares.
\end{itemize}
{\it Author:  }Micha\"el Rao.\\
 Erd\"os formulated two fundamental problems:\\
(1) (1957,1961): Is there arbitrarily long abelian-square-free words on a finite
alphabet?\\
(2) (1961): Is it possible to avoid long squares on a binary alphabet?\\
\ \\
 In 1974,  Entringer, Jackson et Schatz gave a positive answer to the second question \cite{EJS74}. 
In 2002 M\"akel\"a formulated similar questions for the abelian squares or cubes on binary  or ternary alphabets \cite{M02}.
In his talk at WORDS 2015, Rao  presented technics for deciding whether a morphic word avoid abelian and $k$-abelian repetitions: in particular, this allowed him  to prove that long abelian squares are avoidable on a ternary alphabet.
He asked the following questions:
\begin{itemize}
\item
\color{blue}
{\it Question 1.6.15.3:}
\color{black}
Can we avoid long abelian cubes over two letters?
\item
\color{blue}
{\it Question 1.6.15.4:}
\color{black}
How to decide whether a morphic word avoids (long) abelian power?
\item
\color{blue}
{\it Question 1.6.15.5 (due to M\"akel\"a  \cite{EJS74}): }
\color{black}
Let $h$ be the morphism onto $\{0,1,3,4\}^*$ defined  by $h(0)=03$, $h(1)=43$, $h(3)=1$, $h(4)=01$.
Is there a morphism $g:\{0,1,3,4\}^*\longrightarrow \{0,1\}^*$ such that $g(h^\infty(0))$ has no long abelian cubes?
\item
\color{blue}
{\it Question 1.6.15.6: }
\color{black}
Find good heuristics to compute candidates for question 1.6.15.5.
\item
\color{blue}
{\it Question 1.6.15.7: }
\color{black}
Find a morphism simpler than Kur\"anen's one s that avoid abelian square on four letters?
\item
\color{blue}
{\it Question 1.6.15.8:}
\color{black}
What is the minimal $k$ such that one can avoid abelian squares of period at least $k$ over three letters ($2<k<6$)?
\item
\color{blue}
{\it Question 1.6.15.9:}
\color{black}
What is the minimal $k$ such that one can avoid $2$-abelian squares of period at least $k$ over two letters ($2<k<60$)?
\end{itemize}
We refer also the reader to the notions connected to the so-called {\it templates} \cite{ACR04}.
From this point of view, iin 2015 Rao and Rosenfed proved that for any primitive morphism $h$ whose matrix has no eigenvalue of
norm $1$ and any template $t$ it is possible to decide if $h^\infty(a)$ realizes $t$. The following problems may be formulated:
\begin{itemize}
\item
\color{blue}
{\it Question 1.6.15.10:}
\color{black}
Is there a morphism over $5$ letters with two eigenvalues of norm smaller than $1$ and an abelian-square-free fixed point?
\item
\color{blue}
{\it Question 1.6.15.11:}
\color{black}
Is there a morphism on $3$ letters with one eigenvalues of norm smaller than $1$ and an abelian-cube-free fixed point?
\item
\color{blue}
{\it Question 1.6.15.12:}
\color{black}
How to decide if eigenvalues of norm $1$ may be allowed in the result that was mentionned above?
\end{itemize}
\section*{\it 2. Complexity issues}
\addcontentsline{toc}{section}{\it 2. Complexity issues}
In the literature, with a word several notions of complexity can be associated,  the most famous one being certainly the {\it factor complexity}: given a word $w$, this complexity measures the number $p_w(n)$ of different factors of length $n$ occuring in $w$. The famous characterization of Morse-Hedlund for ultimately periodic words led to introduce the infinite  {\it Surmian words} whose complexity is $p_w(n)=n+1$, the best known example of them being certainly the famous {\it Fibonacci word}.
\subsection*{\it 2.1 The recurrence quotient}
\addcontentsline{toc}{subsection}{\it 2.1 The recurrence quotient}
The {\it  recurrence function} has been introduced by Morse and Hedlund: given a  factor $u$, it associates with every non-negative integer $n$ the size $R_u(n)$ of the smallest window that contains every factor of length $n$ of $u$.\\
\ \\
{\it --WORDS 1997}\\
{\it Author: }Julien Cassaigne \cite[3-12]{Issue1997}.\\
The {\it recurrence quotient} is defined as $\rho(u)=\limsup_{n\rightarrow\infty}\frac{R_u(n)}{n}$.\\
For a sturmian sequence of slope $\alpha$, denote the recurrence quotient by $\rho(\alpha)$;
 the spectrum of values of $\rho$ is the set $S$ of the values taken by $\rho(\alpha)$ when $\alpha$ spans $[0,1]\setminus\nbQ$.
\begin{itemize}
\item
\color{blue}
{\it Question 2.1.97.1: }
\color{black}
What is the Hausdorf dimension  (e.g. \cite{HS98}) of $S$ (or that of each of its intervals $S\cap [a,a+1]$)?
\item
\color{blue}
{\it Question 2.1.97.2: }
\color{black}
Draw a study of the recurrence quotients for other families of infinite word that sturmian words, such as words of complexity $2n+1$, or toinfinite words in general.
\end{itemize}
\subsection*{\it 2.2 The ratio $p(n)/n$}
\addcontentsline{toc}{subsection}{\it 2.2 The ratio $p(n)/n$}
Alex Heinis proved that if $p(n)/n$ has a limit, then this limit is either equal to $1$,  or highter than and equal to $2$.\\
\ \\
{\it --WORDS 2001}\\
{\it Author: }Ali Aberkane \cite[31-46]{Issue2001}.\\
By using the so-called Rauzy graphs, in WORDS 2001 the author presents characterizations of the words such that the limit is $1$.
\begin{itemize}
\item
\color{blue}
{\it Question 2.2.01.1:}
\color{black}
Transform the preceding characterization into a characterization using a finite set of substitutions associated with rules governing their composition (i.e. $S$-adic system of representation).
\item
\color{blue}
{\it Question 2.2.01.2: }
\color{black}
Give a characterization of infine words whose complexity satisfies lim$_n~p(n)/n=2$.
\end{itemize}
\subsection*{\it 2.3 The balance function }
\addcontentsline{toc}{subsection}{\it 2.3 The balance function }
~\\
{\it --WORDS 2001}\\
{\it Author:  }Boris Adamczewski \cite[47-75]{Issue2001}.\\
Boris Adamczewski defines the {\it balance function} as 
$\max_{a\in A}~\max_{u,v\in F(w)}\{||u|_a-|v|_a| \}$.
With regard to the so-called {\it primitive substitutions}, the author investigates the connections between the asymptotic behavior of the balance function and the incidence matrix of such a substitution. Moreover, he shows that the Thue-Morse sequence is an example for which the spectrum of the substitution of order two is different of the spectrum of the initial substitution.
\begin{itemize}
\item
\color{blue}
{\it Question 2.3.01.1:}
\color{black}
Give an example of sequence or which the mentionned change of spectrum is really significant for the balance properties. 
\end{itemize}
~\\
{\it --WORDS 2013}\\
{\it Author:  }Julien Cassaigne \cite[1]{Acts2013}.\\
A words is {\it balanced} if for any pairs $(u,v)$, of factors with same length, and for any letter $a$, one have $||u|_a-|v|_a|\leq 1$ (where $|u|_a$ stands for the number of occurrences of the letter $a$ in $u$). A classical characterization  of Sturmian words is that they are the aperiodic $1$-balanced sequences. For Arnoux-rauzy words, whose complexity is $(|A|-1)n+1$, the following question can be formulated (see also \cite{BCS13}):
\begin{itemize}
\item
\color{blue}
{\it Question 2.3.13.1:}
\color{black}
Give characterizations of Arnoux-Rauzy words with a given balance.
\end{itemize}
\subsection*{\it 2.4 The palindromic complexity}
\addcontentsline{toc}{subsection}{\it 2.4 The palindromic complexity}
 The {\it  palindromic complexity} of an infinite word is the function  which counts the number $P(n)$
of different palindromes of length $n$ which occur as factors of this word.\\
\ \\
{\it --WORDS 2005}\\
{\it Authors: }Peter Baláži, Zuzana Mas\'kov\'a and Edita Pelantov\`a \cite[266--275]{Issue2005}.\\
The authors provide an estimate of  $P(n)$ for uniformly recurrent words; denoting by $p(n)$ the classical factor complexity this estimation is based on the equation: $P(n)+P(n+1)=p(n+1)-p(n)+2$.
\begin{itemize}
\item
\color{blue}
{\it Question 2.4.05.1:}
\color{black}
Describe the structure of the Rauzy graphs of words reaching the mentioned upper bound.
\end{itemize}
\section*{\it 3. Factorization of words.  Equations}
\addcontentsline{toc}{section}{\it 3. Factorization of words.  Equations}
Further important information may be obtained by decomposing a word into a convenient sequence of consecutive factors:
$w=w_1\cdots w_n$.
\subsection*{\it 3.1 ${\cal F}$-factorization}
\addcontentsline{toc}{subsection}{\it 3.1 ${\cal F}$-factorization}
The  so-called ${\cal F}$-factorization corresponds to the case where the preceding sequence $(w_1,\cdots, w_n)$ satisfies a given property  ${\cal F}$. Formally, ${\cal F}$ is defined as follows:\\
Let $I=\{1\cdots,k\}$ and $\Sigma$ be two disjoint alphabets. Set ${\cal F}=(L,L_1,\cdots,L_k)$, with $L\subseteq  I^*$ and $L_1,\cdots,L_k\subseteq\Sigma^*$. We say that the sequence of factors $(w_i,\cdots,w_n)$ is a ${\cal F}$-factorization if for all $j\in [1,n]$ we 
have $w_j\in L_{i_j}$ and  ${i_1}\cdots{i_n}\in L$. 
The factorization ${\cal F}$ is {\it regular} ({\it context-free}) if the languages $L,L_1,\cdots, L_k$ are regular (context-free). \\
\ \\
{\it --WORDS 1997}\\
{\it Authors: Juhani Karhum\"aki, Wojciech Plandowski and Wojciech Rytter} \cite[pp. 123--133]{Issue1997}.\\
 Three fundamental properties of ${\cal F}$-factorizations were examined, the so-called {\it completeness}, {\it uniqueness}  and {\it synchronization}.
\begin{itemize}
\item
\color{blue}
{\it Question 3.1.97.1:}
\color{black}
Find efficient algorithms for the polynomial time solvable problems which were discussed in the paper.
\item
\color{blue}
{\it Question 3.1.97.2:}
\color{black}
Given a word, can its minimal and maximal regular ${\cal F}$-factorization, in the sense of the length of the sequence of indices, be found in polynomial time?
\item
\color{blue}
{\it Question 3.1.97.3:}
\color{black}
Could the better algorithms be designed for the  problems discussed in the paper if in regular  ${\cal F}$-factorizations only finite languages are considered?
\item
\color{blue}
{\it Question 3.1.97.4:}
\color{black}
Is the completeness or the uniqueness undecidable if context-free ${\cal F}$-factorizations are given by deterministic automata or by linear context-free grammars?
\item
\color{blue}
{\it Question 3.1.97.5:}
\color{black}
What is the complexity of the problem of determining whether a regular ${\cal F}$-factorization possesses synchronization property if the parameters of the synchronization are not given ? What about this problem for context-free ${\cal F}$-factorizations?
\end{itemize}
\subsection*{\it 3.2 Periodicity}
\addcontentsline{toc}{subsection}{\it 3.2 Periodicity}
With the preceding notation, if for an integer  $n\geq 2$, all the words $w_1,\cdots , w_{n-1}$ are equal, the word $w_n$ being one of their prefixes, we say that the length of $w_1$ is a {\it period} of $w$.
 \\
\ \\
{\it --WORDS 2007}\\
{\it Author: }Kalle Saari \cite[273-279]{Acts2007}.\\
The author proved that  the least period of a non-empty factor of the infinite Fibonacci word is a Fibonacci number.
With regards to  Sturmian words of a given slope, say $\alpha$, he defines the set $\Pi(\alpha)$ as  indicated in the following:\\
Let $[0,1+d_1,d_2,d_3,\cdots]$  the continued fraction expansion of $\alpha$. Set $q_1=q_0=1, q_n=d_nq_{n-1}+q_n~~(n\ge 1)$ and:
$$\Pi(\alpha)=\bigcup_{n\geq 0}\{iq_n+q_{n-1}: i=0,1,\cdots,d_n\}$$
\begin{itemize}
\item
\color{blue}
{\it Conjecture 3.2.07.1:}
\color{black}
Let $t$ denote a Sturmian word with slope $\alpha$. 
If a word is a nonempty factor of $t$, then its least period belongs to  $\Pi(\alpha)$.
\end{itemize}
\subsection*{\it 3.3 Quasiperiodicity}
\addcontentsline{toc}{subsection}{\it 3.3 Quasiperiodicity}
A word $w$ is {\it quasiperiodic} if another word $x$ exists such that any position in $w$ falls within an occurrence of $x$ as a factor of $w$ (informally, $w$ may be completely ``covered" by a set of occurrences of the factor $x$).\\
\ \\
{\it --WORDS 2013}\\
{\it Authors: }Florence Lev\'e and Gwena\"el Richomme \cite[pp. 181--192]{Acts2013}.\\
A morphism is {\it strongly (resp. weakly) quasiperiodic} if it maps any (at least one) non-quasiperiodic word to a quasiperiodic word.
The authors provided algorithms for deciding whether a morphism is strongly quasiperiodic on finite and infinite words.
\begin{itemize}
\item
\color{blue}
{\it Question 3.3.13.1: }
\color{black}
Given a morphism $f$ and a letter $a$ such that $a$ is the initial letter of $f(a)$,  is it decidable that $f^\omega(a)$ is quasiperiodic?\item
\color{blue}
{\it Conjecture 3.3.13.2: }
\color{black}
Let $f$ be an morphism generating a quasiperiodic infinite word.
If $f(a)$ is not a power of $a$ then $f$ weakly quasiperiodic on any infinite word?
\end{itemize}
\subsection*{\it 3.4 Defect effect and independent systems of equations}
\addcontentsline{toc}{subsection}{\it 3.4 Defect effect and independent systems of equations}
The {\it combinatorial rank} of a  set of words $X$, that we denote by $r(X)$, is the smallest number of words needed to express all words of $X$ as products of these words \cite{N92}. 
As a direct consquence of the famous theorem of defect, if $X$ is a not a code (that is, if the words of a finite set $X$ satisfy a nontrivial equation), then  we have $r(X)\le |X|-1$.\\
\ \\
{\it --WORDS 1999}\\
 {\it Authors: }Juhani Karhum\"aki and J\'an Ma\v nuch \cite[pp. 81--97]{Issue1999}.\\
The authors stated the following problems, which are connected to the famous critical factorization theorem \cite[Chap. 8]{Lothaire1}:
\begin{itemize}
\item
\color{blue}
{\it Question 3.4.99.1:}
\color{black}
Let $X$ be a finite set of words, and let $w$ be a non-periodic bi-infinite word. Assume that $w$ possesses $k$ disjoint factorizations, with $k\leq |X|$. Is it true that we have $r(X)\le |X|-k+1$?.
\item
\color{blue}
{\it Question 3.4.99.2:}
\color{black}
Let $X$ be a code, and let $w$ be a bi-infinite word. Assume that for $k\leq |X|$, $w$ possesses $k$ disjoint $X$-factorizations, such that at least one of them is non-periodic. Is it true that we have $r(X)\le  |X|-k+1$?
\item
\color{blue}
{\it Question 3.4.99.3:}
\color{black}
Denote by $p(w)$ the smallest period of a word $w\in\Sigma^+$. Let $X\subseteq \Sigma^+$ satisfying $p(x)< p(w)$ for all $x\in X$. Is it true that $w$ has at most $|X|+1-r(X)$ disjoint $X$-factorizations?
\end{itemize}
{\it --WORDS 2001}\\
{\it Authors: }Tero Harju and Dirk Nowotka \cite[pp. 139--172]{Issue2001}.\\
Defect effect is strongly connected to {\it independent systems of equations}. Given an equation in three variables, say $x,y,z$,  a solution $\alpha$ is {\it non-periodic} if $\alpha(x),\alpha(y),\alpha(z)$ are not powers of the same word \cite[Chapt. 9]{Lothaire1}. A system of equations is {\it independent} if it is not equivalent to any of its proper subsets. An equation is {\it balanced} if the number of occurrences of each variable on the left- and the right-hand side is the same. In their presentation of WORDS 2001, the authors proved that every independent system of equations in three variable with at least two equations and a non-periodic solution consists in balanced equations only. They stated the following question, which was in fact implicitely raised  in 1983 by Culik II and Karhum\"aki
\cite{CIIK83}:
\begin{itemize}
\item
\color{blue}
{\it Question 3.4.01.1:}
\color{black}
Is there an independent system of three equations in three variables with a non-period solution?
\end{itemize}
{\it --WORDS 2005}\\
{\it Authors: }\v St\v ep\'an Holub and Juha Kortelainen \cite[363--372]{Issue2005}.\\
The authors studied the infinite system $(S)$ of words equations with variables $u_k$ ($1\le k\le m$) and $v_{k'}$ ($1\le k'\le n$):
$$\{x_0u_1^ix_1u_2^i x_2\cdots u_m^ix_m= y_0v_1^iy_1v_2^iy_2\cdots v_n^iy_n~:i\ge 0 \} $$
They stated the following questions:
\begin{itemize}
\item
\color{blue}
{\it Question 3.4.05.1:}
\color{black}
Is there a positive integer $k$ such that the system $(S)$ is equivalent to one of its subsystems induced by $q$ equations?
\item
\color{blue}
{\it Question 3.4.05.1:}
\color{black}
Is the system $\{u_1^i=v_1^iv_2^i\cdots v_n^i~~ :i\ge 0\}$ equivalent to one of its subsystems induced by three equations?
\end{itemize}
\subsection*{\it 3.5 The Post Correspondence Problem}
\addcontentsline{toc}{subsection}{\it 3.5 The Post Correspondence Problem}
The famous Post Correspondence Problem ($PCP$ for short)  consists in asking, given two morphisms $h,g$, whether or not the equation $h(x)=g(x)$ has a solution distinct of the empty word.\\
In the most general case, it is well known that this problem is undecidable \cite{P46}. In another hand, many studies were devoted to  special cases of instances (eg. \cite{EKR82}).\\
\newpage
\ \\
{\it --WORDS 2005}\\
{\it Authors: }Vesa Halava, Tero Harju, Juhani Karhum\"aki and Michel Latteux \cite[355-352]{Issue2005}.\\
A morphism $h$ is {\it marked} if for any pair of different letters $a,b$, the initial letters of the words $h(a)$ and $h(b)$ are different.
The words $u,v$ are {\it comparable}, which we denote by  $u\bowtie v$,  is either $u$  is a prefix of $v$ or $v$ is  a prefix of $u$.
With such a notion, special types of instances $(h,g)$ may be defined. In particular $(h, g)$ is called a {\it unique equality continuation instance} if  $h(ua) \bowtie g(ua)$ and $h(ub) \bowtie g(ub)$ imply $h(u) = g(u)$,
for any word
$u$ and any pair of different letters $a, b$.
The authors asked for the two following quesrions:
\begin{itemize}
\item
\color{blue}
{\it Question 3.5.05.1: }
\color{black}
Is PCP decidable for unique equality continuation instances?
\item
\color{blue}
{\it Question 3.5.05.2: }
\color{black}
Is it decidable whether or not an instance of PCP satisfies the property of unique equality continuation instances?
\end{itemize}
%
%

%

\end{document}